# Deep Learning Based Semantic Video Indexing and Retrieval

Anna Podlesnaya, Sergey Podlesnyy
Cinema and Photo Research Institute (NIKFI)
Creative Production Association "Gorky Film Studio"
Moscow, Russia
s.podlesnyy@nikfi.ru

*Abstract*—We share the implementation details and testing results for video retrieval system based exclusively on features extracted by convolutional neural networks. We show that deep learned features might serve as universal signature for semantic content of video useful in many search and retrieval tasks. We further show that graph-based storage structure for video index allows to efficiently retrieving the content with complicated spatial and temporal search queries.

I. INTRODUCTION

In our work we focus on video search or content-based video retrieval for cinematography and television production. Everyday need for footage in TV production consumes much of editors work spent in movie/broadcast archives. Non-fiction movies production also relies on historical and cultural heritage content stored in scattered archives.

Amount of information stored in movie archives is huge. For example in Russian Federation there are National Cinematography Archive storing above 70,000 titles and State TV/Radio Foundation storing above 100,000 titles. Much of their content comprises rare documentary films dated from early XX century to recent days. Putting these materials to modern producers disposal is only possible by means of search techniques.

Well-established methods for searching, navigating, and retrieving broadcast quality video content rely on transcripts obtained from manual annotating, close captioning and/or speech recognition [1]. Recent progress in descriptive audio stream provisioning for the visually impaired has lead to video indexing solutions based on speech recognition of descriptive audio [2].

Video search and data exchange are ruled by international standards, one of most important being MPEG-7 [3]. This standard defines Query Format MPQF in order to provide a standard multimedia query language to unify access to distributed multimedia retrieval systems. Some of query types defined by MPQF are:

- *QueryByMedia* specifies a similarity or exact-match query by example retrieval where the example media can be an image, video, audio or text.

- *QueryByFreeText* specifies a free text retrieval where optionally the focused or to be ignored fields can be declared.

- *SpatialQuery* specifies the retrieval of spatial elements within media objects (e.g., a tree in an image), which may be connected by a specific spatial relation.

- *TemporalQuery* specifies the retrieval of temporal elements within media objects (e.g., a scene in a video), which may be connected by a specific temporal relation.

It is clear that relying on speech recognition techniques is not sufficient to implement the above standards requirements. Querying by media (either by sample image or by sample video clip) is not possible using text-based indexing. Spatial querying would be very much limited as well. One needs to index video by visual content in addition to speech content.

We show in this work that all the above mentioned query types can be implemented using the semantic features extracted from video by deep learning algorithms, namely by convolutional neural networks. Our contribution is (1) presenting a video indexing and retrieval architecture based on unified semantic features and capable to implement MPQF query interface and (2) sharing the results of real world testing.

II. RELATED WORK

There are two possible approaches for video indexing based on visual content: image classification and image description. Image classification approach involves assigning preset tags to every frame, or every key frame, or every scene of a video file. Certain improvements to mere classification task exist including salient objects detection and image segmentation. In case of salient objects detection one tags essentially the bounding boxes found in video frames with preset categories. In case of segmentation one tags free-form image regions. In any case the resulting index includes a set of time codes and categories assigned to corresponding movie parts.

Convolutional neural networks (CNN, e.g. [4, 5, 6]) have recently become de-facto standard in visual classification, segmentation and salient objects detection. For example an architecture described in [5] comprising 19 trainable layers with 144 million parameters achieved 6.8% top-5 error rate at

ILSVRC2014 competition [7]. Authors in [8] expand CNN architecture to video classification by means of temporal pooling and optical flow channel addition to raw frames content.

However CNN are often trained to analyze individual photos that are usually carefully framed and focused for the subject of the image (i.e. the scene) in a clear manner. Videos are typically comprised of "shots" i.e. unit of action in a video filmed without interruption and comprising a single camera view. Within the shots objects may be occluded, blurred, ill positioned (non centered) because the shots are intended for integral perception by the spectators.

Additionally, scene content in videos often varies immensely in appearance, resulting in difficulty in classification of such content. For example, the subject of a video shot may be filmed from different angles and scales within the shot, from panoramic to close-up, causing the subject to appear differently across frames in the shot. Thus, because video often represents wide varieties of content and subjects, even within a particular content type, identification of that content is exceedingly difficult.

Image description approach involves generating natural text annotations based on video frame content. In [9] deep neural network architecture matching image regions with natural language sentence parts is proposed, and multimodal recurrent neural network is proposed that takes images as input and generates their textual descriptions. Using this architecture one can for e.g. generate text descriptions for key frames extracted from video stream and build a searchable index. Since the proposed architecture is capable to generate sentences describing image regions defined by bounding boxes it is possible to apply complex search queries with spatial relations between objects within a key frame.

In [10] text descriptions are generated for video shot i.e. a sequence of frames, using features extracted by CNN (similarly to [9]) and applying soft attention mechanism to generate a description for the shot in the whole.

Image description-based approach has advantages of being friendly for general-purpose search engines like Google or Yandex. However this approach is not efficient for searching by examples as required by MPEG-7 standard. We believe this approach is most promising as accompanying technology for broadcasting quality content retrieval tasks.

Search by example is based on video descriptors. In [11] compact descriptors (28 bit) are obtained by layer-wise training of autoencoder, where every layer is RBM. Compact video descriptors based on oriented histograms are defined in MPEG-7 standard as well [3].

Another aspect of searching by example is due to the fact that current image classifiers typically have a capacity of $10^3$ while reasonable nomenclature of classes suitable for usage in information retrieval amounts to $10^4$ categories of common concepts. In addition, typical search requests include named entities like famous person names, architectural and natural landmarks and brand names (e.g. car models). This makes infeasible the classifiers trained for pre-set known categories only. In [12] an elegant method is proposed involving the HoG features storing in image archive index, and online training of exemplar SVM classifiers based on a set of images (around $10^2$) provided as a template for target concept to be found in the archive. This concept is easy to expand for video archives of course.

### III. VIDEO INDEXING

In this section we describe video indexing architecture.

#### A. Features extraction and film segmenting

We use GoogLeNet network structure [6] as primary source of semantic features extraction. We claim by this work that one-time operation of CNN calculation per frame is enough to build powerful video indexing and retrieval system. For our experiments we use already trained model and image pre-processing protocol described in [6].

First step of video processing pipeline includes features extraction and film segmenting into the shots (see Algorithm 1). In this algorithm, we obtain sub-sampled sequence of movie frames. Sub-sampling period S was chosen as a tradeoff between accuracy and speed, and we found value 320 ms ($1/8^{th}$ frame for standard movie frame rate) to be optimal.

---

**Algorithm 1:** Film segmentation

Input:
    $F = \{f_1, f_2, \ldots f_N\}$ : Video frames sequence
    S : Sampling period (parameter)
    T : Threshold (parameter)

Output:
    $K = \{k_1, k_2, \ldots k_M\}$: Indexes of frames each starting new shot

| | |
|---|---|
| 1: | prev_fv ← Null; K ← {1}; InitFilter(); |
| 2: | **for** i=1 **to** N **with step** S |
| 3: |     fv ← GetFeatureVector($f_i$) |
| 4: |     **if** prev_fv **is not** Null |
| 5: |         d ← Distance(fv, prev_fv) |
| 6: |         df ← Filter(d) |
| 7: |         **if** df > T |
| 8: |             K « i |
| 9: |         **end if** |
| 10: |     **end if** |
| 11: |     prev_fv ← fv |
| 12: | **end for** |

---

Thus, at step 3 we apply *GetFeatureVector* function to the frame to get the feature vector that is used throughout all further operations of indexing and searching. This function includes pre-processing: image re-scaling into 256x256 BGR, selecting single central crop 224x224 and applying the CNN calculation. The function returns an output of the last average-pooling layer of network [6] which has the dimension 1024. In practice, to speed up computations we pack several frames and run calculations in GPU batch mode using *caffe* library [13].

At step 5 we calculate distance between previous and current feature vectors. We are using squared Euclidean distance, however other choices are possible e.g. cosine distance. Fig. 1 shows the typical plot of distance values vs. frame number.

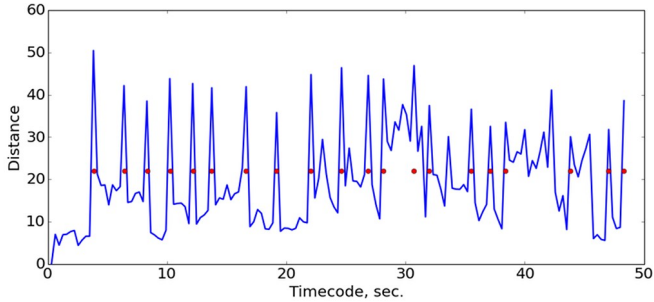

Fig. 1. Distance between neighboring frames feature vectors; red dots indicate shot boundaries detected by Algorithm 1.

Intuitively, since feature vector in CNN being the source for SOFTMAX classifier contains semantic information of the frame, we expect that frames with similar content would have close feature vectors. The shot in video is a sequence of frames filmed at single camera view thus it would normally contain similar objects and background in all frames. Thus a shot boundary happens where frame content differs dramatically from the previous shot, and feature vectors differ substantially. In cinema shot boundaries are often made soft with dilution or darkening effects. However CNN has shown to be robust to illumination condition of images, so darkening effects usually are treated well. Dilution effect where objects from previous shot are blended with objects from new shot produce spikes in the plot similar to Fig. 1, and are easy to filter out.

Filtering operation is performed at step 6. We are using simple low-pass filter e.g. convolution of 4-window of last distance values with vector [0.1, 0.1, 0.1, 0.99]. Then at step 7 we check if filtered value of vector distance exceeds a threshold value, and add frame number to shot boundary list if it exceeds.

Sample results of shots detection are shown in Fig. 2. In order to evaluate this algorithm we compared shot boundaries with I-Frame positions in MPEG-4 encoded movie. I-Frames are used by MPEG-4 codec as base frames stored without compression, while consecutive frames are encoded as difference values from latest I-Frame. Thus I-Frame are good candidates to shot boundaries because they are inserted into the video stream specifically when scene changes dramatically and difference encoding becomes not feasible.

We obtained precision 0.935 and recall 0.860 considering MPEG-4 I-Frames positions as ground truth while varying sampling period S (see Algorithm 1). We considered shot boundary as true if its index was within 5 frames from ground truth. Naturally, as sampling period grows we loose some shot boundaries hence recall decreases. However precision stays almost at the same level justifying the fact that we are using feature vector encoding frame semantics.

Relatively low recall is explained with the fact that I-Frames are inserted by MPEG-4 codec in order to minimize reconstruction error in video stream. Therefore it may insert numerous semantically similar key frames having just a small visual difference. Algorithm 1 considers the shot by its semantic contents and produces fewer shot boundaries.

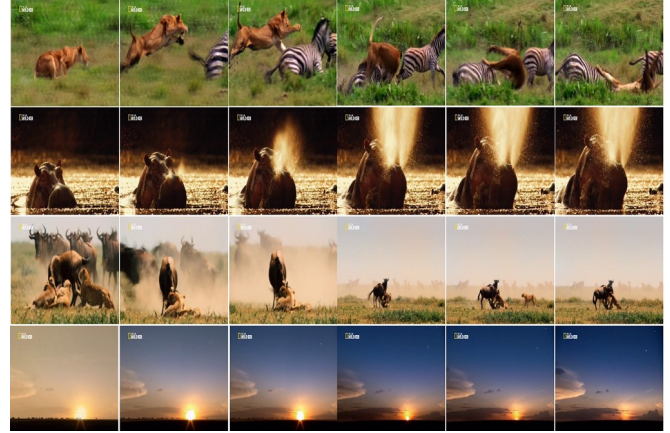

Fig. 2. Example shots detected in "The great Serengeti", National Geographic, 2011 movie fragment.

As a side product of Algorithm 1 we store feature vectors and classification vectors (CNN output) for every frame into a distributed key-value storage (Apache Cassandra). This is the only time when we apply CNN calculation. Technically it may mean that from this point we do not need GPU for efficient functioning of video indexing and video retrieval. The rest operation may be performed in inexpensive cluster or cloud-based infrastructure with CPU-only server nodes.

*B. Graph-oriented indexing*

We introduce a graph structure for building a video index. This is partially due to the fact that trained model [6] that we use predicts categories within ImageNet contest framework [7] which uses Wordnet [14] lexical database. This lexical database is essentially a graph representation of words (synsets) connected with linguistic relations such as "hypernum", "part holonym" etc. This opens wide possibilities for video retrieval by description e.g. by a query for videos where an object being part of some general category is required. We represent the Wordnet lexical database with graph

$$G_{WORDNET} = (N_{NOUNS}, E_{LEXICAL\_RELATIONS}) \quad (1)$$

where N denotes nodes, E - edges.

The main unit of graph-based representation of videos is a shot. As will be shown below, CNN classifier more accurately categorizes shots than single frames. From user experience point of view, retrieving shots is natural in case of video searching.

In our experiments we used CNN trained for ILSVRC2014 competition [7]. It was trained for 1000 categories, majority of which were dogs and flowers species as well as many other animals. This is biased from what we may expect in categorizing common videos. Therefore we chose BBC Natural World (2006) series of 102 movies, each approx. 45 minutes long for evaluating the proposed system. For practical use it will be enough to train classifier with common objects in order to remove this bias to natural history.

Our evaluation of per-frame classification by top-5 score using 1056 random video frames labeled manually yielded accuracy 0.36±0.11. This is much lower than 0.93 reported in [6] but of course this is due to the fact that ImageNet dataset is closed in a sense that every image does have correct tags belonging to 1000 categories known to the classifier.

We then performed classification vectors temporal pooling. Concretely we pooled *min(10, <number of frames in the shot>)* classification vectors and compared the accuracy for average pooling and max pooling. The difference between pooling methods was vanishing, and accuracy rose to 0.46±0.23. If we further consider Wordnet lexical hierarchy and treat as correct classifications one-step hypernum from the category predicted by CNN (e.g. CNN predicted *cheetah* while true category is *leopard*, both share same hypernum *big_cat*) the resulting accuracy was 0.53±0.23. Therefore we choose to index shots by average pooling the classification vectors and to provide an option for the retrieval of shots using hypernum to the queried keyword.

In section 3A we simplistically presented video processing as classifying every frame with single CNN. In reality we could apply numerous classifiers e.g. place classifier, faces detector and classifier, salient objects detector and classifier. Having applied all that classifiers we might obtain numerous tags for a frame. Moreover, these tags may also have structure e.g. if we detect two salient objects we may consider spatial relationship between them: which object is atop or right to the second one. Therefore it becomes natural to represent a film as a graph:

$$G_{FILM} = (N, E), \quad (2)$$
$$N = \{N_{SHOTS}, N_{TAGS}\},$$
$$E = \{E_{CATEGORIES}, E_{PLACES}, E_{FACES}, E_{SALIENT\_OBJ}, E_{SPATIAL}\}$$

It is clear that we may link $G_{FILM}$ with $G_{WORDNET}$ by matching $N_{TAGS}$ with $N_{NOUNS}$. Fig. 3 illustrates possible graph representation of a film comprising two shots.

We used Neo4j graph-oriented database for video index for its excellent implementation of Cypher query language [15]. The expressive querying of Cypher is inspired by a number of different approaches and established practices from SQL, SPARQL, Haskell and Python. Its pattern matching syntax looks like ASCII art for graphs, which will be shown in IV.

## IV. VIDEO RETRIEVAL

In this section we describe an implementation of video retrieval modes required by MPEG-7 standard.

### A. Searching by Structured Queries

Basic keywords-based search in our graph index can be implemented with Cypher statement (3). It accounts for minimum confidence level of shot tags, and sorts the search results by shot duration descending.

$$\text{MATCH (s:Shot) - [c:Category] ->} \quad (3)$$
$$\text{(w:Wordnet \{synset: "zebra"\})}$$
$$\text{WHERE c.weight > 0.1}$$
$$\text{RETURN s ORDER BY s.duration DESC}$$

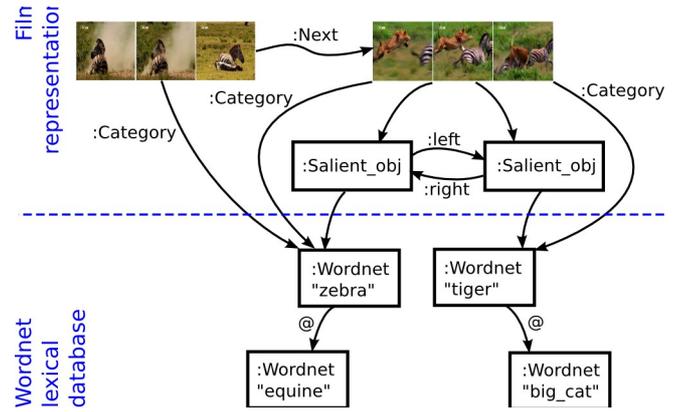

Fig. 3. Graph representation of a film.

Basic Cypher syntax rules denote graph nodes in round brackets and edges in square brackets. Thus query (3) matches nodes of type Shot: $N_{SHOTS}$, see (2) linked to nodes of type Wordnet: $N_{NOUNS}$, see (1) having synset *zebra* with edge of type Category having weight greater than 0.1. Edge of type Category corresponds to $E_{CATEGORIES}$ in (2). Neo4j provides indexing by nodes/edges attributes, therefore performance of this query is quite good. In our test archive storing 99,505 shots the query with additional LIMIT/SKIP clause took approx. 40 ms. In our tests the average precision of queries by 40 random keywords from ImageNet contest categories nomenclature was 0.84±0.25. We could not afford recall evaluation because of a lack of labeled video content, but in information retrieval precision is more important from user point of view: when user searches for *zebras* they definitely don't want to see *fish* in search result (authors are aware of *zebrafish* existence).

It is easy to extend (3) to search for combinations of keywords, as well as logical combinations (AND, OR, NOT).

One way to improve recall is to include synonyms and/or hypernums into the search query. Graph representation of Wordnet lexical database allows easy solution in our video index by query (4). Here we at first match the hypernum of *cheetah* (which is a *big_cat*), and then match all shots having a path to the *big_cat* node. Such type of query limited by 10 results was executed in approx. 40 ms in our tests.

$$\text{MATCH (w:Wordnet \{synset: "cheetah"\}) -} \quad (4)$$
$$\text{[lr:Lexical\_rel] -> (big\_cats:Wordnet)}$$
$$\text{MATCH (s:Shot) - [c:Category] -> () --> (big\_cats)}$$
$$\text{WHERE c.weight > 0.1 and lr.symbol = "@"}$$
$$\text{RETURN s ORDER BY s.duration DESC}$$

Also we might build a query matching the video shots having certain structure. Let's find videos having a *lion* to the left from a *zebra* (5).

$$\text{MATCH (s:Shot) --> (zebra\_obj:Salient\_obj) -->} \quad (5)$$
$$\text{(w:Wordnet \{synset: "zebra"\})}$$
$$\text{MATCH (s) --> (lion\_obj:Salient\_obj) -->}$$
$$\text{(w:Wordnet \{synset: "lion"\})}$$

```
MATCH (zebra_obj) - [:Left] -> (lion_obj)
RETURN s ORDER BY s.duration DESC
```

*B. Searching by Sample Video*

Video retrieval by sample clip is important in content production (finding footage in archives) and in duplicates finding (for legal purposes and for archives deduplication). In our setting the sample video is limited to a single shot discussed above, and the goal is to find semantically close shots. This differs from many existing solutions based on e.g. HSV histograms or SIFT/SURF descriptors.

We found that feature vector fv $\in \mathcal{R}^{1024}$ extracted in Algorithm 1 contains enough semantic information for retrieving video shots having similar content with the sample clip. A brute force solution involves comparing distance between sample clip feature vector and every other shot's feature vector with some threshold, and including the shots having smaller distance to the sample into the search results. We compared Euclidean distance and cosine distance metrics of vector distance and selected the cosine distance as preferred one (6).

$$d = 1 - dot(x, y) \qquad (6)$$

Where x - sample clip feature vector, y - other clip feature vector.

In order to improve the performance we apply Wordnet hierarchy to limit the scope of shots to check. Concretely we select one or two hypernums of the categories of sample clip by query (4). Only the shots matching this condition are cycled through vector distance check. Thus we look at the shots having similar lexical content and select the closest ones by feature vector distance. This results in retrieving the relevant shots by terms that are hard to formalize, see Fig. 5. Figure 4(a) shows the results of a search by keyword *elephant*. From these results a user have chosen a sample shot where a herd of elephants, a lake and forest are filmed. Searching by this sample retrieved a number of shots having these characteristics proving that one image is better than hundreds of words - see Fig. 4(b).

Average precision of search by video was 0.86. We evaluated precision by searching by a keyword and then searching by one of resulted shots with cosine distance threshold 0.3. A human expert counted true/false positives. We used 42 keywords for this evaluation. Figure 4(c) shows the distribution of precision values measured by different keywords.

*C. Searching by Sample Images*

In order to extend possibilities for video retrieval beyond the scope of pre-set nomenclature of categories we explored on-line training of linear classifiers over feature vectors extracted by CNN.

In order to train classifier we obtained around 100 positive samples by querying images search engine like Yandex or Google. E.g. we queried Yandex for *steamboat* and chose 100 first search results. We scaled every image to 256x256 BGR pixels and applied CNN [6] to both straight and horizontally flipped central patch 224x224 px. We thus obtained 200 feature vectors from the output of layer "pool5/7x7_s1".

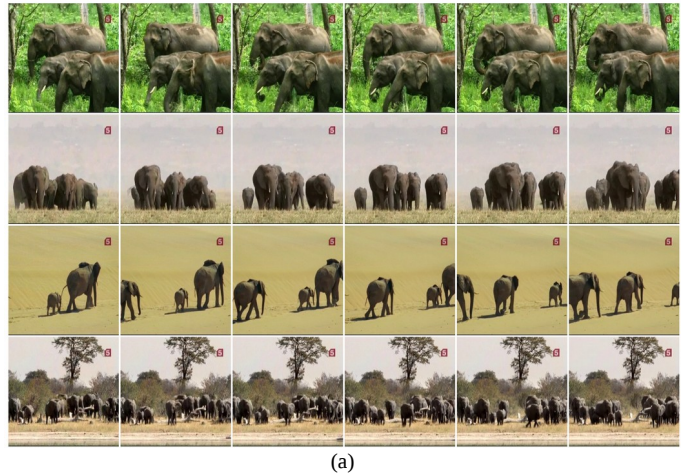

(a)

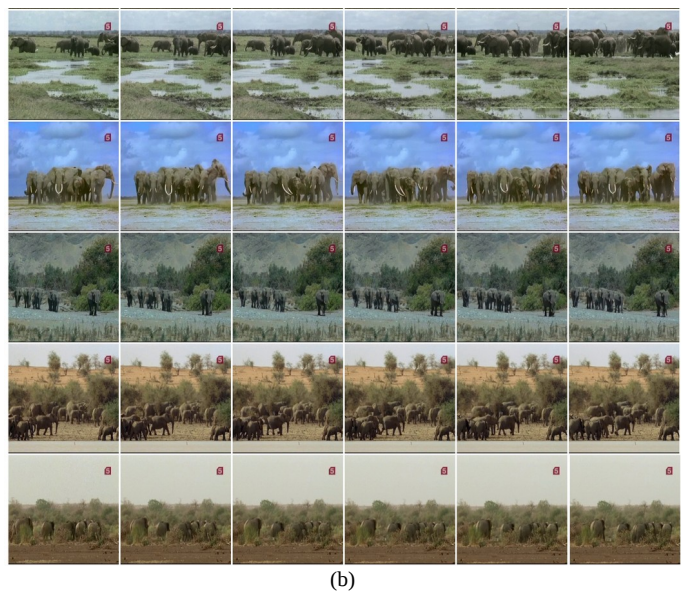

(b)

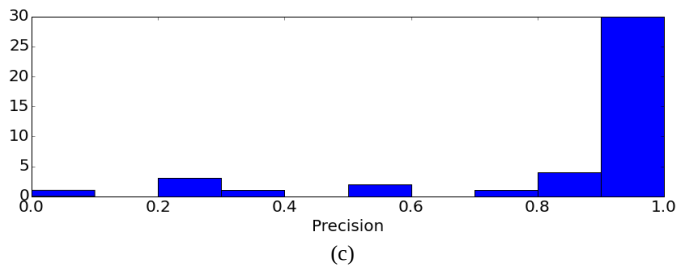

(c)

Fig. 4. Search by example use case: (a) search results by keyword "elephant"; (b) results of searching by sample clip – the last row of Fig. 5(a); (c) histogram of precision values measured for different keywords.

For negative samples we randomly selected 25,000 shots from our test archive, and averaged feature vectors of first K frames of each shot, $K = min(10, N_{FRAMES\_IN\_SHOT})$.

For online training we randomly shuffled positive and negative samples and applied Vowpal Wabbit [16] to train a logistic regression classifier. The following parameters differed from default values: positive sample weight 200, epochs number 3, learning rate 0.5. Training took less than a second in standard Intel-based PC.

A brute force solution involves applying the trained classifier to every shot's feature vector, and including the shots having positive classification into the search results.

Average precision of search by sample images was 0.64. We evaluated precision by obtaining sample images from Yandex by a random keyword and then searching our test archive by 100 sample images. A human expert counted true/false positives. We used 13 keywords for this evaluation. Figure 5(a) shows some of the sample images from Yandex, Fig. 5(b) shows some video shots retrieved from our test archive, Fig. 5(c) shows the distribution of precision values measured by different search requests.

## V. CONCLUSION

We showed in this work that feature vector fv $\in \mathcal{R}^{1024}$ extracted by CNN [6] contains enough semantic information for segmenting raw video into shots with 0.92 precision; retrieving video shots by keywords with 0.84 precision; retrieving videos by sample video clip with 0.86 precision and retrieving videos by online learning with 0.64 precision. All that is needed for indexing is a single pass of feature vector extraction and storing into the database. This is the only time when expensive GPU-enabled hardware is needed. All video retrieval operations may run in commodity servers e.g. in cloud-based setting.

However more efforts are necessary to increase the performance of samples-based video retrieval. While lexical pruning of search space helps to limit the scope for brute force algorithm it scales linearly with the data amount. We plan to explore several approaches for lowering the feature vector dimensionality in order to search in log time scale, e.g. random projections and compact binary descriptors.

ACKNOWLEDGMENTS

This work was funded by Russian Federation Ministry of Culture contract No. 2214-01-41/06-15.

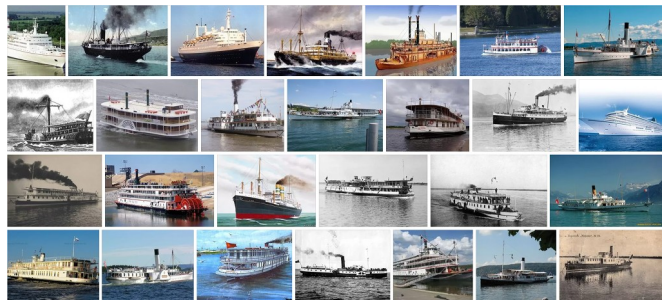

(a)

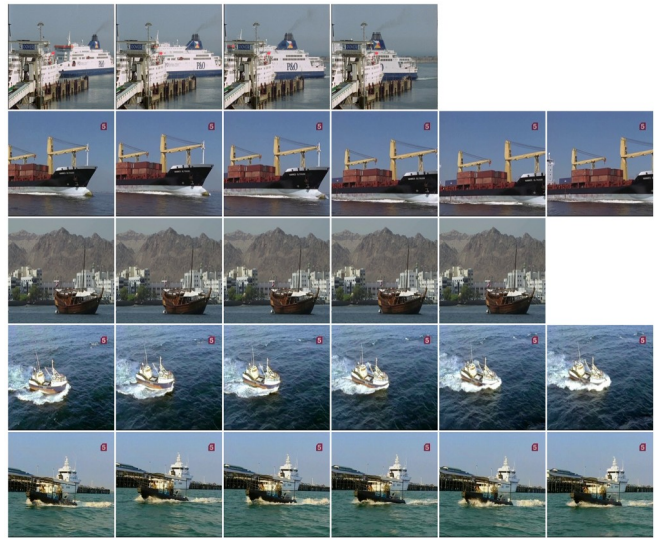

(b)

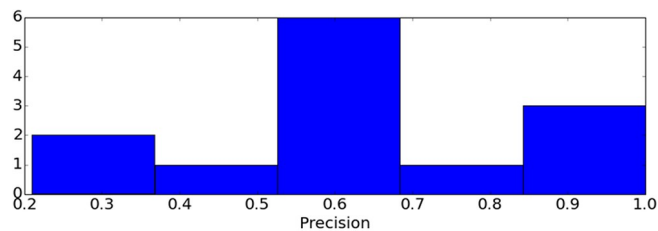

(c)

Fig. 5. Search by sample images use case: (a) sample images obtained from Yandex by query "steamboat"; (b) some video clips retrieved from test archive; (c) histogram of precision values by various requests for sample images.